\documentclass[aps,preprint]{revtex4}%
\usepackage{amsfonts}
\usepackage{amsmath}
\usepackage{amssymb}
\usepackage{graphicx}%
\providecommand{\U}[1]{\protect\rule{.1in}{.1in}}
\begin{document}
\title[ ]{The Contrasting Roles of Planck's Constant in Classical and Quantum Theories}
\author{Timothy H. Boyer}
\affiliation{Department of Physics, City College of the City University of New York, New
York, New York 10031}
\keywords{}
\pacs{}

\begin{abstract}
We trace the historical appearance of Planck's constant in physics, and we
note that initially the constant did not appear in connection with quanta.
\ Furthermore, we emphasize that Planck's constant can appear in both
classical and quantum theories. \ In both theories, Planck's constant sets the
scale of atomic phenomena. \ However, the roles played in the foundations of
the theories are sharply different. \ In quantum theory, Planck's constant is
crucial to the structure of the theory. \ On the other hand, in classical
electrodynamics, Planck's constant is optional, since it appears only as the
scale factor for the (homogeneous) source-free contribution to the general
solution of Maxwell's equations. \ Since classical electrodynamics can be
solved while taking the homogenous source-free contribution in the solution as
zero or non-zero, there are naturally two different theories of classical
electrodynamics, one in which Planck's constant is taken as zero and one where
it is taken as non-zero. \ The textbooks of classical electromagnetism present
only the version in which Planck's constant is taken to vanish.

\end{abstract}
\maketitle

\section{Introduction}

\textquotedblleft Planck's constant $h$ is a `quantum
constant\textquotedblright' is what I am told by my students.
\ \textquotedblleft Planck's constant is not allowed in a classical
theory\textquotedblright\ is the view of many physicists. \ I believe that
these views misunderstand the role of physical constants and, in particular,
the role of Planck's constant $h$ within classical and quantum theories. \ It
is noteworthy that despite the current orthodox view which restricts the
constant $h$ to quantum theory, Planck searched for many years for a way to
fit his constant $h$ into classical electrodynamics, and felt he could find no
place. \ Late in life, Planck\cite{SA} acknowledged that his colleagues felt
that his long futile search \textquotedblleft bordered on a
tragedy.\textquotedblright\ \ However, Planck's constant $h$ indeed has a
natural place in classical electrodynamics. \ Yet even today, most physicists
are unaware of this natural place, and some even wish to suppress information
regarding this role. \ In this article, we review the historical\cite{history}
appearance of Planck's constant, and then emphasize its contrasting roles in
classical and quantum theories. \ 

\section{Physical Constants as Scales for Physical Phenomena}

Physical constants appear in connection with measurements of the scale of
natural phenomena. \ Thus, for example, Cavendish's constant $G$ appeared
first in connection with gravitational forces using the Newtonian theory of
gravity. \ At the end of the 18th century, Cavendish's experiment using a
torsion balance in combination with Newton's theory provided the information
needed for an accurate evaluation of the constant. \ However, because physical
constants refer to natural phenomena and are not the exclusive domain of any
one theory, Cavendish's constant $G$ also reappears in Einstein's 20th-century
general-relativistic description of gravitational phenomena. \ 

In a similar fashion, Planck's constant $h$ sets the scale of electromagnetic
phenomena at the atomic level. \ It was first evaluated in 1899 in connection
with a fit of experimental data to Wien's theoretical suggestion for the
spectrum of blackbody radiation. \ Subsequently, the constant has reappeared
both in quantum theory and in classical electrodynamics. \ In its role as a
scale, the constant $h$ can appear in any theory which attempts to explain
aspects of atomic physics.

\section{Physical Constants of the 19th Century}

The 19th century saw developments in theories involving a number of physical
constants which are still in use today. \ The kinetic theory of gases involves
the gas constant $R$ and Avogadro's number $N_{A}$. \ The unification of
electricity and magnetism in Maxwell's equations involves an implicit or
explicit (depending on the units) appearance of the speed of light in vacuum
$c.$ \ Measurements of blackbody radiation led to the introduction of new
physical constants. \ Thus the appearance of the Stefan-Boltzmann law
$\mathcal{U}_{T}=a_{S}T^{4}V$ for the thermal energy $\mathcal{U}_{T}$ of
radiation in a cavity of volume $V$ at temperature $T$ introduced a new
constant $a_{S},$ Stefan's constant, in 1879. Today this constant is
reexpressed in terms of later constants as $a_{S}=\pi^{2}k_{B}^{4}%
/(15\hbar^{3}c^{3}).$ \ In the 1890s, careful experimental work on the
spectrum of blackbody radiation led to Wien's theoretical suggestion for the
blackbody radiation spectrum with its two constants, (labeled here as $\alpha$
and $\beta),$%
\begin{equation}
\rho_{W}(\nu,T)=\alpha\nu^{3}\exp[-\beta\nu/T].\label{Wien}%
\end{equation}
Also at the end of the century, the measurement of the ratio of charge to mass
for cathode rays led to new constants involving the charge and mass of the electron.

\section{Appearance of Planck's Constant}

Planck's great interest in thermodynamics led him to consider the equilibrium
of electric dipole oscillators when located\ in random classical radiation.
\ Planck found that the random radiation spectrum could be connected to the
average energy $U(\nu,T)$ of an electric dipole oscillator of natural
frequency $\nu$ as $\rho(\nu,T)=(8\pi\nu^{2}/c^{3})U(\nu,T)$.~\ \ Introducing
Wien's suggestion $\rho_{W}$ in (\ref{Wien}) for the spectral form, Planck
found for the average energy of an oscillator%
\begin{equation}
U_{W}(\nu,T)=h\nu\exp[-\beta\nu/T].
\end{equation}
Here is where Planck's constant $h$\ first appeared in physics. \ At the
meeting of the Prussian Academy of Sciences on May 18, 1899, Planck
reported\cite{May1899} the value $\beta$ as $\beta=0.4818\times10^{-10}$
sec$\cdot$K$^{o}$ and the value of the constant $h$ as $h=6.88510^{-27}$
erg-sec. \ Thus initially, Planck's constant appeared as a numerical fit to
the experimental blackbody data when using theoretical ideas associated with
Wien's proposed form for the radiation spectrum. \ Planck's constant $h$ had
nothing to do with quanta in its first appearance in physics.

In the middle of the year 1900, experimentalists Rubens and Kurlbaum found
that their measurements of the blackbody radiation spectrum departed from the
form suggested by Wien in Eq. (\ref{Wien}). \ It was found that at low
frequencies (long wavelengths) the spectrum $\rho(\nu,T)$ seem to be
proportional to $\nu^{2}T.$ \ Planck learned of the new experimental results,
and, using ideas of energy and entropy for a dipole oscillator, introduced a
simple interpolation between the newly-suggested low-frequency form and the
well-established Wien high-frequency form. \ His interpolation gave him the
Planck radiation form corresponding to a dipole oscillator energy
\begin{equation}
U_{P}(\nu,T)=\frac{h\nu}{\exp[\beta\nu/T]-1}. \label{Planck}%
\end{equation}
Planck reported\cite{Oct1900} this suggested blackbody radiation spectrum to
the German Physical Society on October 19, 1900. \ The Planck spectrum
(\ref{Planck}) involved exactly the same constants as appeared in his work
when starting from Wien's spectrum. \ The experimentalists confirmed that
Planck's new suggested spectrum was an excellent fit to the data. \ Once again
Planck's constant $h$ appeared as a parameter in a fit to experimental data
when working with an assumed theoretical form. \ There was still no suggestion
of quanta in connection with Planck's constant $h.$

Planck still needed a theoretical justification for his new spectral form.
\ Although Planck had hoped initially that his dipole oscillators would act as
black particles and would bring random radiation into the equilibrium
blackbody spectrum, he had come to realize that his small linear oscillators
would not change the frequency spectrum of the random radiation; the incident
and the scattered radiation were at the same frequency. \ In late 1900 in an
\textquotedblleft act of desperation,\textquotedblright\ Planck turned to
Boltzmann's statistical work which he had earlier \textquotedblleft vehemently
rejected.\textquotedblright\ \ It was in connection with the use of
statistical ideas that Planck found that the constant $\beta$ in the radiation
spectra could be rewritten as $\beta=h/k_{B}$ where $k_{B}=R/N_{A},$ where $R$
and $N_{A}$ were the constants which had already appeared in the kinetic
theory of gases. \ Indeed, it was Planck who introduced Boltzmann's constant
$k_{B}$ into physics. \ Boltzmann had always stated that entropy $S$ was
proportional to the logarithm of probability $W$ without ever giving the
constant of proportionality. \ Planck introduced the equality $S=k_{B}\ln W.$
\ Also, in his calculations of the probability, Planck departed from
Boltzmann's procedures in retaining the energy connection $\mathcal{E}=h\nu$
without taking the expected limit $\mathcal{E\rightarrow}0.$ \ Only by
avoiding the limit, could Planck recover his radiation spectrum. \ It was here
in retaining $\mathcal{E}=h\nu,$ rather than taking Boltzmann's limit, that
the association of Planck's constant $h$\ with quanta first appeared. \ 

\section{Planck's Constant in Current Quantum Theory}

Although Planck's constant did not originally appear in connection with
quantum theory, the constant became a useful scale factor when dealing with
the photoelectric effect, specific heats of solids, and the Bohr atom. \ The
theoretical context for all these phenomena was quantum theory. \ Quantum
theory developed steadily during the first third of the 20th century. \ Today
quantum theory is regarded as the theory which gives valid results for
phenomena at the atomic level. \ 

Our current textbooks emphsize that quantum theory incorporates Planck's
constant $h$ into the essential aspects of the theory involving non-commuting
operators. \ Thus position and momentum are associated with operators
satisfying the commutation relation $[\widehat{x},\widehat{p}_{x}%
]=ih/(2\pi)=i\hbar.$ \ The non-commutativity is associate with a non-zero
value of Planck's constant $h$ and disappears when $h$ is taken to zero.
\ Associated with the non-commuting operators is the zero-point energy
$U=(1/2)h\nu_{0}$ of a harmonic oscillator of natural frequency $\nu_{0}.$
\ The oscillator zero-point energy vanishes along with a vanishing value for
$h.$

\section{Planck's Constant in Classical Electrodynamics}

The natural place for Planck's constant within classical electrodynamics is as
the scale factor of the source-free contribution to the general solution of
Maxwell's equations. \ Maxwell's equations for the electromagnetic fields
$\mathbf{E}(\mathbf{r},t)$ and $\mathbf{B}(\mathbf{r,}t)$ can be rewritten in
terms of the potentials $\Phi(\mathbf{r},t)$ and $\mathbf{A}(\mathbf{r},t)$
where $\mathbf{E}=-\nabla\Phi-(1/c)\partial\mathbf{A}/\partial t$ and
$\mathbf{B}=\nabla\times\mathbf{A.}$ \ In the Lorenz gauge, Maxwell's
equations for the potentials become wave equations with sources in the charge
density $\rho(\mathbf{r},t)$ and current density $\mathbf{J}(\mathbf{r},t),$%
\begin{equation}
\left(  \nabla^{2}-\frac{1}{c^{2}}\frac{\partial^{2}}{\partial t^{2}}\right)
\Phi(\mathbf{r},t)=-4\pi\rho(\mathbf{r},t), \label{scalar}%
\end{equation}%
\begin{equation}
\left(  \nabla^{2}-\frac{1}{c^{2}}\frac{\partial^{2}}{\partial t^{2}}\right)
\mathbf{A}(\mathbf{r},t)=-4\pi\frac{\mathbf{J}(\mathbf{r},t)}{c}.
\label{vector}%
\end{equation}
The general solutions of these differential equations in all spacetime with
outgoing wave boundary conditions are
\begin{equation}
\Phi(\mathbf{r},t)=\Phi^{in}(\mathbf{r},t)+%
%TCIMACRO{\tint }%
%BeginExpansion
{\textstyle\int}
%EndExpansion
d^{3}r^{\prime}%
%TCIMACRO{\tint }%
%BeginExpansion
{\textstyle\int}
%EndExpansion
dt^{\prime}\frac{\delta(t-t^{\prime}-|\mathbf{r-r}^{\prime}|/c)}%
{|\mathbf{r-r}^{\prime}|}\rho(\mathbf{r}^{\prime},t^{\prime}),
\end{equation}

\begin{equation}
\mathbf{A}(\mathbf{r},t)=\mathbf{A}^{in}(\mathbf{r},t)+%
%TCIMACRO{\tint }%
%BeginExpansion
{\textstyle\int}
%EndExpansion
d^{3}r^{\prime}%
%TCIMACRO{\tint }%
%BeginExpansion
{\textstyle\int}
%EndExpansion
dt^{\prime}\frac{\delta(t-t^{\prime}-|\mathbf{r-r}^{\prime}|/c)}%
{|\mathbf{r-r}^{\prime}|}\frac{\mathbf{J}(\mathbf{r}^{\prime},t^{\prime})}{c},
\end{equation}
where $\Phi^{in}(\mathbf{r},t)$ and $\mathbf{A}^{in}(\mathbf{r},t)$ are the
(homogeneous) source-free contributions to the general solutions of Eqs.
(\ref{scalar}) and (\ref{vector}). \ It is universally accepted that these are
the accurate solutions. \ However, all the textbooks\cite{texts} of
electromagnetism omit the source-free contributions $\Phi^{in}(\mathbf{r},t)$
and $\mathbf{A}^{in}(\mathbf{r},t),$ and present only the contributions due to
the charge and currents sources $\rho(\mathbf{r},t)$ and $\mathbf{J}%
(\mathbf{r},t).$ \ 

Within a laboratory situation, the experimenter's sources correspond to
$\rho(\mathbf{r},t)$ and $\mathbf{J}(\mathbf{r},t)$ while the source-free
terms $\Phi^{in}(\mathbf{r},t)$ and $\mathbf{A}^{in}(\mathbf{r},t)$ correspond
to contributions the experimenter does not control which are present when the
experimenter enters his laboratory. \ The terms $\Phi^{in}(\mathbf{r},t)$ and
$\mathbf{A}^{in}(\mathbf{r},t)$ might correspond to radio waves coming from a
nearby broadcasting station, or might correspond to thermal radiation from the
walls of the laboratory. \ In a shielded laboratory held at zero temperature,
there is still random classical radiation present which can be measured using
the Casimir force\cite{Casimir} between two parallel conducting plates.
\ Intrepreted within classical electromagnetic theory, this Casimir force
responds to all the classical radiation surrounding the conducting parallel
plates, and it is found experimentally\cite{exp} that this force does not go
to zero as the temperature goes to zero. \ The Casimir force, interpreted
within classical electromagnetic theory, indicates that there is classical
electromagnetic zero-point radiation present throughout spacetime. \ The
zero-point spectrum is Lorentz invariant with a scale set by Planck's constant
$h.$ \ Thus in a shielded laboratory at zero-temperature, the vector potential
$\mathbf{A}(\mathbf{r},t)$ should be written as
\begin{align*}
\mathbf{A}(\mathbf{r},t) &  =%
%TCIMACRO{\tsum _{\lambda=1}^{2}}%
%BeginExpansion
{\textstyle\sum_{\lambda=1}^{2}}
%EndExpansion%
%TCIMACRO{\tint }%
%BeginExpansion
{\textstyle\int}
%EndExpansion
d^{3}k\widehat{\epsilon}(\mathbf{k},\lambda)\left(  \frac{h}{4\pi^{3}\omega
}\right)  ^{1/2}\sin\left[  \mathbf{k}\cdot\mathbf{r-}\omega t+\theta
(\mathbf{k},\lambda)\right]  \\
&  +%
%TCIMACRO{\tint }%
%BeginExpansion
{\textstyle\int}
%EndExpansion
d^{3}r^{\prime}%
%TCIMACRO{\tint }%
%BeginExpansion
{\textstyle\int}
%EndExpansion
dt^{\prime}\frac{\delta(t-t^{\prime}-|\mathbf{r}-\mathbf{r}^{\prime}%
|/c)}{|\mathbf{r}-\mathbf{r}^{\prime}|}\frac{\mathbf{J}(\mathbf{r}^{\prime
},t^{\prime})}{c},
\end{align*}
including a source-free contribution which is a Lorentz-invariant spectrum of
plane waves with random phases $\theta(\mathbf{k},\lambda)$ and with Planck's
constant $h$\ setting the scale.\cite{Breview}

As indicated here, there is a natural place for Planck's constant $h$ within
classical electromagnetic theory. \ Once this zero-point radiation is present
in the theory, it causes zero-point energy for dipole oscillators, and it can
be used to explain Casimir forces, van der Waals forces, oscillator specific
heats, diamagnetism, the blackbody radiation spectrum, and the absence of
\textquotedblleft atomic collapse.\textquotedblright\cite{any} \ 

\section{Contrasting Roles for Planck's Constant}

Planck's constant $h$\ sets the scale for atomic phenomena. \ However, the
constant plays contrasting roles in classical and quantum theories. \ Within
quantum theory, Planck's constant $h$ is embedded in the essential aspects of
the theory. \ If one sets Planck's constant to zero, then the quantum
character of the theory disappears; the non-commuting operators become simply
commuting c-numbers, and the zero-point energy of a harmonic oscillator drops
to zero.

On the other hand, within classical electrodynamics, Planck's constant does
not appear in Maxwell's fundamental differential equations. \ Rather, Planck's
constant appears only in the (homogeneous) source-free contribution to the
general solution of Maxwell's equations. \ Thus classical electrodynamics can
exist in two natural forms. \ In one form Planck's constant is taken as
non-zero, and one can explain a number of natural phenomena at the atomic
level. \ In the other form, Planck's constant is taken to vanish. \ It is only
this second form which appears in the textbooks of electromagnetism and modern
physics. \ However, even this form is quite sufficient to account for our
macroscopic electromagnetic technology.

\section{Comments on Planck's Constant in Classical Theory}

Planck was a \textquotedblleft reluctant revolutionary\textquotedblright\ who
tried to conserve as much of 19th century physics as possible. \ Late in life
Planck wrote\cite{SA} in his scientific autobiography, \textquotedblleft My
futile attempts to fit the elementary quantum of action somehow into the
classical theory continued for a number of years, and they cost me a great
deal of effort. \ Many of my colleagues saw in this something bordering on a
tragedy.\textquotedblright\ \ Within his lifetime, there seems to have been no
recognition of a natural place for Planck's constant $h$ within classical
electrodynamics. \ It was only in the 1960s, beginning with careful work by
Marshall,\cite{Marshall} that it became clear that Planck's constant $h$ could
be incorporated in an natural way within classical electrodynamics.

Despite the realization that Planck's constant is associated with atomic
phenomena and that at least some of atomic phenomena can be described within
either classical or quantum theory, there seems great reluctance on the part
of some physicists to acknowledge the possibility of Planck's constant
appearing within classical theory. \ One referee wrote the following
justification\cite{personal} in rejecting the idea of Planck's constant
appearing within classical electromagnetism: \textquotedblleft But as a
pedagogical matter, doesn't it muddy the distinction between classical and
quantum physics? \ The traditional dividing line may be in some aspects
arbitrary, but at least it is clear ($\hbar\implies$
quantum).\textquotedblright\ \ This referee clearly misunderstands the nature
of physical constants and seems willing to sacrifice scientific accuracy to
pedagogical simplicity.

I believe that many physicists would prefer truth to convenient pedagogy.
\ Certainly the introduction of Planck's constant as the scale factor for
source-free classical zero-point radiation expands the range of phenomena
described by classical electromagnetic theory.\cite{any} \ The recognition
that Planck's constant has a natural place within classical theory may provide
a broadening perspective beyond a confining quantum orthodoxy.\ \ In any case,
one suspects that Planck would have been pleased to find that there is indeed
a natural role for his constant $h$ within classical electromagnetic theory.

(revised October 1, 2017)


\begin{thebibliography}{99}                                                                                               %


\bibitem {SA}M. Planck, \textit{Scientific Autobiography and Other Papers,
}translated by F. Gaynor (Philosophical Library, New York 1949), p. 35.

\bibitem {history}The historical information is taken from the following
sources: T. S. Kuhn, \textit{Black-Body Theory and the Quantum Discontinuity
1894-1912} (Oxford U. Press, New York 1978); \ A. Hermann (translated by C. W.
Nash) \textit{The Genesis of Quantum Theory (1899-1913)} (MIT Press,
Cambridge, MA 1971); M. J. Klein, \textquotedblleft Planck, Entropy, and
Quanta 1901-1906,\textquotedblright\ The Natural Philosopher I, pp.. 81-108
(Blaisdell Publishing Co. New York 1963); \ M. J. Klein, \textquotedblleft
Thermodynamics and Quanta in Planck's Work,\textquotedblright\ in
\textit{History of Physics} edited by S. R. Weart and M. Philips (AIP, New
York 1985), pp. 294-302.

\bibitem {May1899}M. Planck, S.-B. Preuss. Akad. Wiss. (1899), p. 440.

\bibitem {Oct1900}M. Planck, Verh. d. Deutsch. Phys. Ges. \textbf{2}, 202
(1900). \ 

\bibitem {texts}See for example, D. J. Griffiths, \textit{Introduction to
Electrodynamics 4th ed.} (Pearson, New York 2013), section 10.2.1 Eq. (10.26);
or L. Eyges, \textit{The Classical Electrodynamic Field }(Cambridge University
Press 1972), p. 186, Eqs. (11.45) and (11.46); or J. D. Jackson,
\textit{Classical Electrodynamics 3rd ed.} (John Wiley \& Sons, New York,
1999), 3rd ed., p. 246, Eq. (6.48); or A. Zangwill, \textit{Modern
Electrodynamics} (Cambridge U. Press, 2013), p. 724-725, Eqs. (20.57) and
(20.58); or A. Garg, \textit{Classical Electromagnetism in a Nutshell}
(Princeton U. Press, Princeton, NJ 08450, 2012), p. 204, Eqs.(54.16) and (54.17).

\bibitem {Casimir}H. B. G. Casimir, \textquotedblleft On the attraction
between two perfectly conducting plates,\textquotedblright\ Proc. Ned. Akad.
Wetenschap. \textbf{51}, 793-795 (1948).

\bibitem {exp}M. J. Sparnaay, \textquotedblleft Measurement of the attractive
forces between flat plates,\textquotedblright\ Physica (Amsterdam)
\textbf{24}, 751-764 (1958); S. K. Lamoreaux, \textquotedblleft Demonstration
of the Casimir force in the 0.6 to 6 $\mu$m range,\textquotedblright\ Phys.
Rev. Lett. \textbf{78}, 5-8 (1997): \textbf{81}, 5475-5476 (1998); U.
Mohideen, \textquotedblleft Precision measurement of the Casimir force from
0.1 to 0.9 $\mu$m,\textquotedblright\ \textit{ibid.} \textbf{81}, 4549-4552
(1998); H. B. Chan, V. A. Aksyuk, R. N. Kleinman, D. J. Bishop, and F.
Capasso, \textquotedblleft Quantum mechanical actuation of
microelectromechanical systems by the Casimir force,\textquotedblright%
\ Science \textbf{291}, 1941-1944 (2001): G. Bressi, G. Caarugno, R. Onofrio,
and G. Ruoso, \textquotedblleft Measurement of the Casimir force between
parallel metallic surfaces,\textquotedblright\ Phys. Rev. Lett. \textbf{88},
041804(4) (2002).

\bibitem {Breview}T. H. Boyer, \textquotedblleft Random electrodynamics: The
theory of classical electrodynamics with classical electromagnetic zero-point
radiation,\textquotedblright\ Phys. Rev. D \textbf{11}, 790-808 (1975).

\bibitem {any}T. H. Boyer, \textquotedblleft Any classical description of
nature requires classical electromagnetic zero-point
radiation,\textquotedblright\ Am. J. Phys. \textbf{79}, 1163-1167 (2011).
\ See also, D. C. Cole and Y. Zou, "Quantum Mechanical Ground State of
Hydrogen Obtained from Classical Electrodynamics," Phys. Lett. A \textbf{317},
14-20 (2003). \ A review of the work on classical electromagnetic zero-point
radiation up to 1996 is provided by L. de la Pena and A. M. Cetto, \textit{The
Quantum Dice - An Introduction to Stochastic Electrodynamics} (Kluwer
Academic, Dordrecht 1996).

\bibitem {Marshall}T. W. Marshall, \textquotedblleft Random
electrodynamics,\textquotedblright\ Proc. R. Soc. \textbf{A276}, 475-491
(1963). \ T. W. Marshall, \textquotedblleft Statistical
Electrodynamics,\textquotedblright\ Proc. Camb. Phil. Soc. \textbf{61},
537-546 (1965).

\bibitem {personal}Author's personal correspondence. \ \ \ 
\end{thebibliography}
\end{document}